# Domain Wall Conductivity in Oxygen Deficient Multiferroic YMnO$_3$ Single Crystals

Y. Du[1], X. L. Wang,[1]* D. P. Chen,[1] S. X. Dou,[1] Z. X. Cheng,[1] M. Higgins,[2] G. Wallace,[2] J. Y. Wang[3]

[1] *Institute for Superconducting and Electronic Materials, AIIM, University of Wollongong, North Wollongong, New South Wales 2522, Australia*

[2] *Intelligent Polymer Research Institute, AIIM, University of Wollongong, North Wollongong, New South Wales 2522, Australia*

[3] *Institute of Crystal Growth, Shandong University, Jinan, Shandong 250100, P.R. China*

**Abstract**

The transport properties of domain walls in oxygen deficient multiferroic YMnO$_3$ single crystals have been probed using conductive atomic force microscopy and piezoresponse force microscopy. Domain walls exhibit significantly enhanced conductance after being poled in electric fields, possibly induced by oxygen vacancy ordering at domain walls. The electronic conduction can be understood by the Schottky emission and Fowler-Nordheim tunnelling mechanisms. Our results show that the domain wall conductance can be modulated through band structure engineering by manipulating ordered oxygen vacancies in the poling fields.

*To whom correspondence should be addressed: xiaolin@uow.edu.au

Manipulation of electronic states in complex oxides is of significant importance and promising for the next generation electronics applications.[1] In particular, ferroic materials, which possess spontaneous polarization, magnetization, and strain, are regarded as the candidate materials for electronic state modulation, due to the possibility of controlling the various degrees of freedom at domain walls (DWs), where the ferroic order parameters interact with electronic states and crystal structures.[2-7] Recently, oxygen vacancy induced dynamic conductance has been observed at DWs in multiferroic systems, such as $BiFeO_3$.[8, 9] The DW conductance can be tuned through control of the polarization and structural distortion at DWs. However, the origin of this defect-induced DW conductance is still not fully understood. $YMnO_3$ is a well-known multiferroic compound, in which structural anti-phase boundaries, and ferroelectric and magnetic DWs interlock, forming multiferroic DWs.[8,9] The electric dipoles are driven by Y $d^0$ behavior in $YMnO_3$, in which the Y $4d^0$ state is strongly hybridized with the O $2p$ ones. The distortion of $Y^{3+}$ ions results in a structural trimerization that induces an improper ferroelectricity.[12] Since structural anti-phase boundaries in $YMnO_3$ are also antiferromagnetic domain boundaries below its Néel temperature ($T_N$ = 80 K), the ferroelectricity, magnetization, and structure couple together at the multiferroic DWs. In terms of unique structure-driven multiferroic coupling in $YMnO_3$, the topological defects, *e.g*. DWs, are expected to control the transport properties through modulating the electronic states, which is the underlying basis of its applications.

In this work, a significant enhancement of local conductance at DWs in $YMnO_3$ single crystals was observed by piezoresponse force microscopy and conductive atomic force microscopy. The conductance is tunable by controlling oxygen vacancy ordering at DWs. The enhanced transport properties can be understood by the Schottky emission and Fowler-Nordheim mechanisms, as a consequence of ordered oxygen vacancies at DWs. The results provide a possibility of manipulating electronic states in complex oxides through design and by controlling the electronic structure and chemical pressure at DWs.

$YMnO_3$ single crystals were grown by the floating zone method in an infrared radiation furnace equipped with four 1000 W halogen lamps (Crystal System Inc.). The feed rods for the crystal growth were prepared by the conventional solid-state method. Stoichiometric $Y_2O_3$ and $Mn_2O_3$ were mixed and pressed into pellets at 200 MPa. The mixtures were sintered at 1150 °C for 24 h, and then were sintered at 1350 °C for 20 h. YMnO$_3$ and YMnO$_{3-\delta}$ crystals were grown in both air (P$_{O2}$ = 0.1 MPa) and argon (P$_{Ar}$ = 0.4 MPa) atmospheres, respectively. The domain dynamics in Ar-grown YMnO$_{3-\delta}$ crystals was studied by applying an external electric field through the top and bottom electrodes coated on both sides of the sample. Poling was carried out in electric fields up to 80 kV/cm along $c$ axis. After poling, electrodes were removed by a target surfacing system (Leica EM TXP). For piezoresponse force microscopy (PFM, Asylum Research MFP-3D) and conductive atomic force microscopy (CAFM, Asylum Research MFP-3D) measurements, plate-like crystals with a platelet area of 1 mm$^2$ and a thickness of 50 µm along the $c$-axis were prepared by an ion beam cutter system (EM TIC020, Leica). Pt/Ir coated Si cantilevers (tip radius ≈ 28 nm) with force constant of 2.8 N/m were used in both PFM and CAFM.

Fig. 1 shows in-plane PFM images of the $a$-$b$ surface of YMnO$_3$ single crystals. The scanning areas for air-grown YMnO$_3$ and Ar-grown YMnO$_{3-\delta}$ crystals are 5 × 5 µm$^2$ and 12 × 12 µm$^2$. "Cloverleaf" 180° domain structures are observed in the air-grown YMnO$_3$ crystal, as shown in Fig. 1(a), which is consistent with the report in Ref. [10]. It shows six domains with downward $P_\downarrow$ (bright areas) and upward $P_\uparrow$ (dark areas) where the polarizations form domain vortices that join at a clamping point. The domain size is 0.5-2 µm. Because the ferroelectric DWs interlock with the structural DWs, the ferroelectric domain structure is induced by a structural trimerization that has different phases rotated by 60° with respect to each other. Fig. 1(b) shows the domain structure of the Ar-grown YMnO$_{3-\delta}$ crystal. Instead of the domain vortices, it is found that, in most cases, two $P_\downarrow$ domains are connected via a thin bridge but separated from the third $P_\downarrow$ domain. Ferroelectric domains with straight DWs have been observed. The domain size in YMnO$_{3-\delta}$ is several micrometers, which is larger than that observed in air-grown YMnO$_3$ crystal. After being poled under an electric field (80 kV/cm), a quasi-stripe domain pattern was observed in Ar-grown YMnO$_{3-\delta}$ by PFM, as shown in Fig. 1(c). In YMnO$_3$ crystals, strong anisotropic hybridization of the 4$d$-2$p$ (Y$^{3+}$-O$^{2-}$) states has been determined by X-ray absorption spectroscopy along the $c$ axis.[12] The ferroelectricity is induced by the opposite polarizations between two distorted YO$_8$ units with a 1:2 ratio. The oxygen vacancies in YMnO$_3$ are expected to influence the Y$^{3+}$-O$^{2-}$ hybridization, which, in turn, engineers the domain structures. In stoichiometric YMnO$_3$ crystals, the oxygen vacancies can be formed in diluted amounts and are considered as uncorrelated randomly distributed point defects that scarcely influence the domain configurations. Nevertheless, the oxygen vacancies in YMnO$_{3-\delta}$ tend to become ordered as vacancy clusters due to their high concentration, initially forming one-dimensional chains, and above a certain vacancy density, ordering in planes of specific geometries.[13-15] The DWs provide a stable energy state which leads to accumulation of ordered oxygen vacancies. These ordered defects will induce the YO$_8$ distortions and reduce the strength of the 4$d$-2$p$ (Y$^{3+}$-O$^{2-}$) hybridization, leading to the local wall structures that arise from the interaction of vacancy clusters with DWs. As a result, straight DWs, instead of twisted ones, were tailored according to the ordered-oxygen-vacancies defects. However, the investigation of scanning tunnelling microscopy is necessary for the direct evidence of the possible oxygen vacancy ordering at DWs.

The local conductivity of Ar-grown YMnO$_{3-\delta}$ crystal was probed by CAFM. Local currents were measured in the regions including DWs, as shown in Fig. 2. The crystal was poled in an external electric field up to 80 kV/cm. YMnO$_{3-\delta}$ crystal exhibits a homogenous insulating behaviour before poling, as shown in Fig. 2(a). As different poling fields were applied, it was found that DWs in YMnO$_{3-\delta}$ became conductive and showed clear contrast to the domains in the CAFM current image (Fig. 2(b)). The conductance of DWs was observed to be hundreds femtoamperes in the 40 kV/cm-poled sample and was further increased to more than 30 picoamperes in the 80 kV/cm-poled sample. The current line profiles of CAFM images indicate that the domains, in contrast to DWs, remain insulating in the poled crystals. The width of the conductive DWs was 50-100 nm. The conductance along the DWs was found to be heterogeneous. This conductive behaviour is completely different to that in stoichiometric YMnO$_3$ crystal, in which the DWs are homogeneous and insulating.[8]

Local $I$-$V$ measurements were further carried out by CAFM in order to understand the local conductance mechanism at DWs in high-field-poled YMnO$_{3-\delta}$ crystals. As shown in Fig. 3(a), $I$-$V$ curves were acquired from three locations, which are on domains, on the edges of DWs, and at DWs, on the $a$-$b$ surface. During the measurements, the conductive AFM tip was stepped in a perpendicular direction with the same force load on the sample. The ferroelectric domain exhibits constant conductance of several femtoamperes with various biases applied, which indicates that the domain is insulating. In contrast, *I-V* curves for DWs show a linear conductance in the low bias region, and have a significant transition to a curvature when high bias is applied. This interface transport behaviour follows the Schottky emission mechanism, which can be expressed as

$$I = \lambda A_0 T^2 \exp\left[\frac{e}{k_B T}\left(\frac{eE}{4\pi\varepsilon\varepsilon_0}\right)^{1/2} - \frac{\phi}{k_B T}\right]$$

where $A_0$ is the Richardson constant, $k_B$ is the Boltzmann constant, $\lambda$ is a correction factor, $\varepsilon_0$ is the dielectric constant, and $\Phi$ is the Schottky barrier. By plotting current vs. $V^{1/2}$ on the log scale for *I-V* curves at DWs and DW edges (Fig. 3(b)), the data follow straight lines in the high bias region. The Schottky barrier height, $\Delta\phi$, of DW edges is calculated to be larger than that of DWs by ~0.29 eV, as shown in Fig. 4. *I-V* data was also plotted in Fowler-Nordheim coordinates,[8] as shown in Fig. 3(b) inset. Linearity in the region of high bias and current shows that Fowler-Nordheim (FN) tunnelling also exist when the higher bias was appied on the tip. It, therefore, indicates that both Schottky emission and FN tunnelling mechanisms are in principle compatible with the data in the limited bias range. We noticed that the similar phenomenon has been recently reported for La-doped BiFeO$_3$ multiferroic thin films.[8] It should be noted that the observed very low DWs conductance in our YMnO$_{3-\delta}$ single crystal could be limited by the tip-surface interface.

A schematic of the band structures at different positions (domain, edge of DW, and DW) is shown in Fig. 4. The enhanced DW conductance in the high bias region is induced by oxygen vacancy activation in an electric field. The oxygen vacancy clusters are regarded as mobile positive charges, which can migrate to thermodynamically stable positions under electric fields.[16] The creation of ferroelectric DWs in ferroelectrics introduces stable defect states on the DWs and frustrated defects on the edges of DWs.[17,18] In the poling process, the ionized oxygen vacancies in YMnO$_{3-\delta}$ move toward DWs in order to lower their energy states. Since YMnO$_3$ is a *p*-type multiferroic semiconductor, the ionized vacancies regarded as charge carriers lower the Schottky barriers by lifting and increasing the valence band. As a result, conductance of the DWs is higher than that of the domains. In addition, the different conductivity across the DWs is likely due to difference in the density of oxygen vacancies.

In conclusion, tunable electronic conductivity has been observed at domain walls in oxygen-deficient $YMnO_3$. The transport properties follow the Schottky emission and Fowler-Nordheim tunnelling mechanisms. After the poling process, the oxygen vacancies are thermally activated and might accumulate at DWs, resulting in changes in the local band structure, which in turn enhances the conductance of DWs. These results provide a crucial step towards electronic applications based on electronic state manipulation through design and control of the electronic structure and chemical pressure at DWs.

This work is supported by Australian Research Council (ARC) through Discovery Projects DP0987190 and DP0558753.

Figure captions:

FIG. 1. PFM images of the *a-b* surface of (a) air-grown $YMnO_3$ and (b) Ar-grown $YMnO_{3-\delta}$ single crystals. Bright and dark areas correspond to the ferroelectric domains with $P_\downarrow$ and $P_\uparrow$, respectively. (c) PFM image of the *a-b* surface of $YMnO_{3-\delta}$ after poling in an electric field of 80 kV/cm. The scale bars represents 1 µm in (a) and 2 µm in (b) and (c).

FIG. 2. CAFM current images of $YMnO_{3-\delta}$ single crystal: (a) without poling; (b) poled under external field of 40 kV; and (c) poled under external field of 80 kV. (d), (e), and (f) are the current profiles corresponding to the red lines marked in (a), (b), and (c), respectively. The scale bars in (a), (b), and (c) represent 0.5 µm, 0.25 µm, and 0.5 µm, respectively.

FIG. 3. (a) *I-V* curves measured by CAFM on $YMnO_{3-\delta}$ *a-b* surface. The inset shows the corresponding measurement locations. (b) Plot of the current against $V^{1/2}$ on the log scale, which indicates a typical Schottky emission mechanism when high bias is applied. The inset shows the *I-V* data plotted in Fowler-Nordheim coordinate.

FIG. 4. (a) Schematic diagrams of interfacial band structure between conductive tip and domains ($V_{tip} = 0$). DWs. $\Delta\Phi_\downarrow$, $\Delta\Phi_\uparrow$, $\Delta\Phi_{DW}$ is the Schottky barriers corresponding to the $P_\downarrow$, $P_\uparrow$ and DWs, respectively. CB, EF, and VB represent for conduction band, Fermi level, and valence band, respectively. (b) Schematic local band structure for the domain wall in oxygen deficient $YMnO_{3-\delta}$ single crystal.

**FIG. 1**

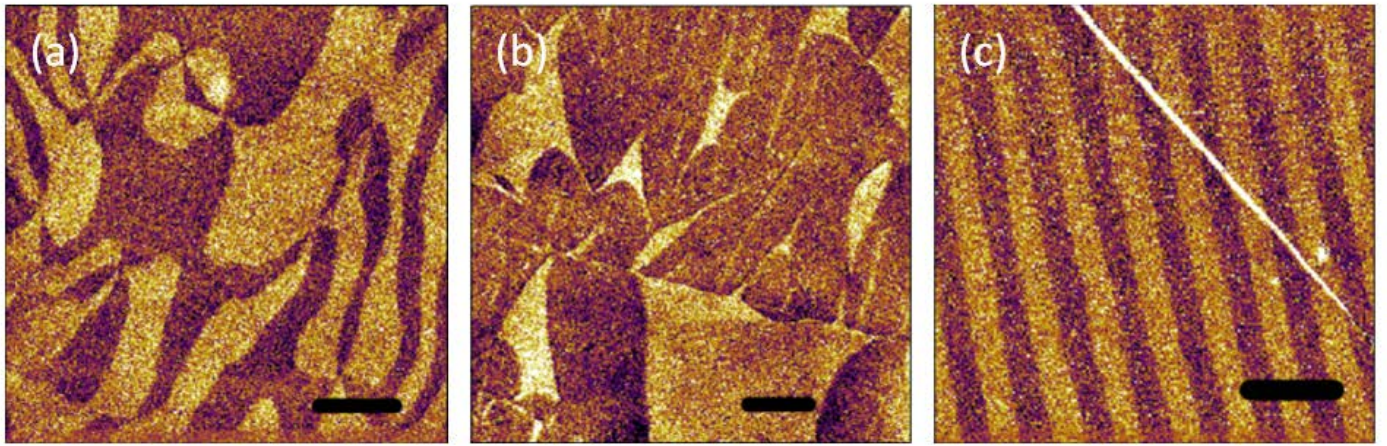

**FIG. 2**

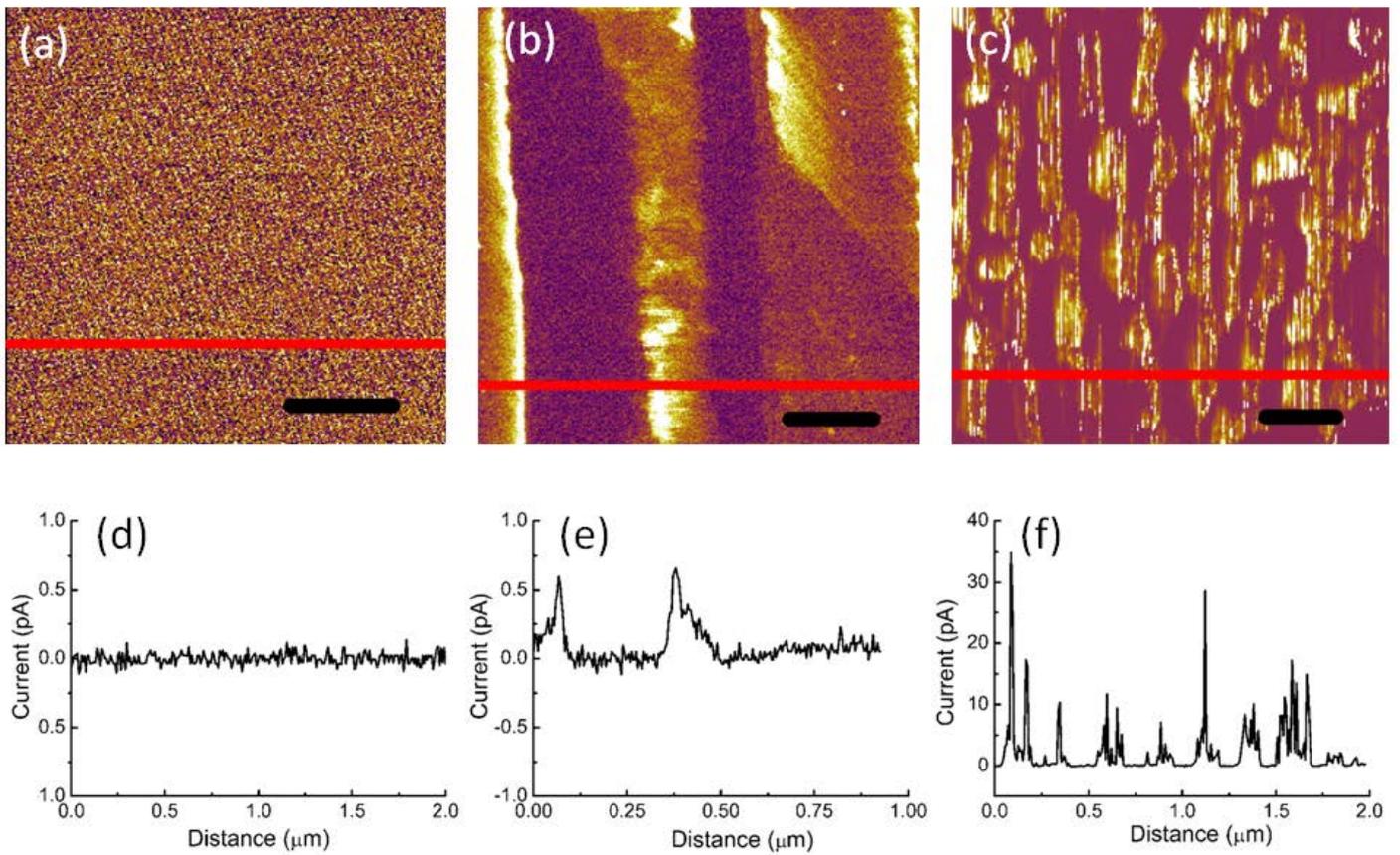

**FIG. 3**

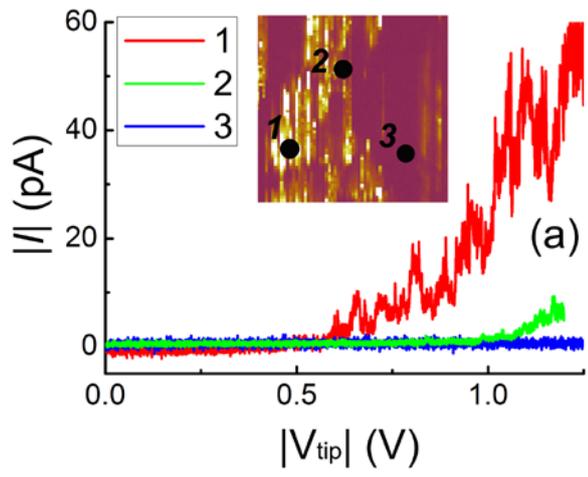 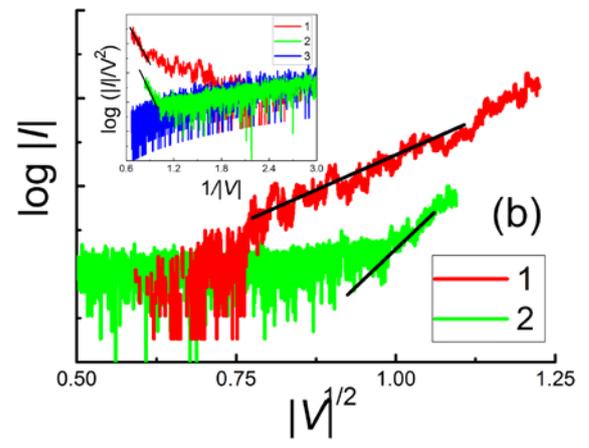

**FIG. 4**

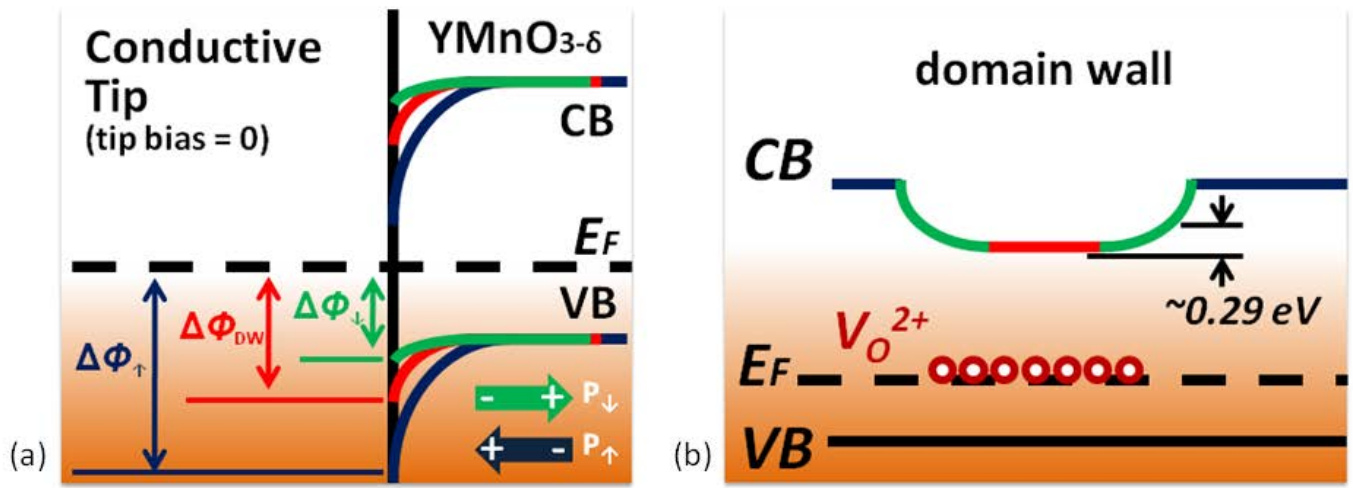